\documentclass[prb,showpacs]{revtex4}
\usepackage{graphicx,psfrag,amssymb,amsmath,feynmf}

\begin{document}
\title{Impurity scattering in unconventional density waves}
\author{Bal\'azs D\'ora}
\affiliation{Department of Physics, Technical University of Budapest, H-1521 Budapest, Hungary}
\author{Attila Virosztek}
\affiliation{Department of Physics, Technical University of Budapest, H-1521 Budapest, Hungary}
\affiliation{Research Institute for Solid State Physics and Optics, P.O.Box
49, H-1525 Budapest, Hungary}
\author{Kazumi Maki}
\affiliation{Max Planck Institute for the Physics of Complex Systems, 
N\"othnitzer Str. 38, D-01187, Dresden, Germany}

\date{\today}

\begin{abstract}
We have investigated the effect of nonmagnetic impurities on the quasi-one-dimensional 
unconventional density wave (UDW) ground state. The thermodynamics were 
found to be close to those of a d-wave superconductor in the Born limit.
Four different optical conductivity curves were found depending on the 
direction of the applied electric field and on the wavevector dependence 
of the gap. 
\end{abstract}

\pacs{75.30.Fv, 71.45.Lr, 72.15.Eb, 72.15.Nj}

\maketitle

\section{Introduction}
Recently a number of papers has been published investigating the different
properties of unconventional density waves (UDW) under various conditions. The common
feature of these systems is the zero average of the gap on the Fermi surface,
resulting in the absence of any periodic modulation of the charge or spin
density. Clearly this property makes UDW a very likely candidate for those
systems in which clear thermodynamic signals of a phase transition
are detected without any obvious order parameter\cite{Sudip}. From this 
the notion "hidden-order" follows naturally. 

Unconventional density wave formation is possible in a large variety of systems.
In the quasi-one-dimensional case, which is the natural occurence of density
waves\cite{Gruner},
 we have investigated the basic properties of
unconventional spin and charge density waves (USDW, UCDW)\cite{nagycikk} and the related threshold 
electric field
with\cite{tesla} and without magnetic field\cite{rapid}. UCDW turned out to be
relevant 
in the explanation of response of low
temperature phase of quasi-one-dimensional $\alpha$-(BEDT-TTF)$_2$KHg(SCN)$_4$ salts. 
In two dimensional systems, the different unconventional phases were
elaborated by Ozaki\cite{Ozaki}. Among them, the d-density wave scenario which is a
special case of UDW (orbital antiferromagnet), was proposed
recently to
describe the famous pseudogap phase of high $T_c$ superconductors\cite{nayak}.
Since the original proposal, several works have been
published in which the properties of d-density waves were studied with the aim
of testing the validity of the model by comparing to experimental 
data (see Ref. \onlinecite{carbotte} and the references therein). 
Also the ground state of certain heavy fermion materials were suspected to be USDW
\cite{IO,roma} which would simply explain the unsolved problem of 
micromagnetism. 
In the presence of magnetic field, the orbital antiferromagnet\cite{Ners1} and the spin 
nematic state\cite{Ners2} were discussed as well in two dimensions.
In three dimensional systems, the pseudogap phase of the transition metal
oxides have attracted significant attention and the staggered flux state was
mentioned in the context of the possible explanations\cite{3dflux}.

In this paper we extend our earlier analysis\cite{nagycikk} on 
pure unconventional density waves to the presence of nonmagnetic impurities. 
Impurities are treated in the Born scattering limit since it works very
well for conventional DW. Since the Fermi surface of quasi-one-dimensional
systems mainly consists of two separate sheets, two different scattering
processes should be taken into account: forward and backward scattering during
which an electron remains on the same or moves to the other Fermi sheet,
respectively.
The thermodynamics are found to be similar to those of a d-wave 
superconductor
in the Born limit. Among the transport properties the quasiparticle part of
the optical conductivity is evaluated. In the chain direction the phason
couples strongly to the electromagnetic field, giving rise to massive
collective modes in this direction. On the other hand, for electric fields
applied perpendicular to the conducting chain, the conductivity shows only
Fermi liquid renormalization, and our description is valid under 
these circumstances.

\section{Formalism}
To start with, we consider the Hamiltonian of interacting electrons:
\begin{equation}
 H=\sum_{\bf k,\sigma}\xi({\bf k})a_{\bf k,\sigma}^{+}a_{\bf
 k,\sigma}+ \frac{1}{2V}
   \sum_{\begin{array}{c}
          {\bf k,k^\prime,q} \\
          \sigma,\sigma^\prime
         \end{array}} \tilde{V}({\bf k,k^\prime,q})a_{\bf k+q,\sigma}^{+}
         a_{\bf k,\sigma}a_{\bf k^\prime-q,\sigma^\prime}^{+}a_{\bf
         k^\prime,\sigma^\prime} ,
\label{hamilton}
\end{equation}
where $a_{\bf k,\sigma}^{+}$ and $a_{\bf k,\sigma}$ are, respectively,
the creation and annihilation operators of an electron of momentum $\bf k$ and
spin $\sigma$. $V$ is the volume of the sample.  
Our system is based on an orthogonal lattice, with lattice constants $a,b,c$
toward direction $x,y,z$. The system is anisotropic, the quasi-one-dimensional
direction is the $x$ axis.
The kinetic-energy spectrum of the Hamiltonian is:
\begin{equation}
\xi({\bf k})=-2t_{a}\cos(k_{x}a)-2t_{b}\cos(k_{y}b)-2t_{c}\cos(k_{z}c)-\mu.
\end{equation}
In the second term of Eq. (\ref{hamilton}) we consider the interaction between on site and
nearest neighbor electrons as in Ref. \cite{nagycikk}. By moving from Bloch
space to Wannier space, the Wannier function is well localized, leading to
a significant dependence of the interaction matrix element on the incoming
electron momenta $\bf k$ and $\bf k^\prime$. Its antisymmetrized 
(therefore spin dependent) version\cite{klasszikus}
is given by
\begin{gather}
\frac{N}{V}\tilde{V}({\bf k,k^\prime,q,\sigma,\sigma^\prime})=\delta_{-\sigma,
\sigma^\prime}(U+\sum_{i}(2V_{i}\cos
q_{i}\delta_{i}+2J_{i}\cos(k_{i}-k_{i}^\prime+q_{i})\delta_{i}+ \nonumber \\
+2Re(F_{i}e^{i (k_{i}^{'}+k_{i})\delta_{i}})+
2Re(C_{i}(e^{i k_{i}\delta_{i}}+e^{i k_{i}^\prime\delta_{i}}+
 e^{i (k_{i}^\prime-q_{i})\delta_{i}}+e^{i (k_{i}+q_{i})\delta_{i}}))))+ \nonumber \\
  +\delta_{\sigma,\sigma^\prime}\sum_{i}(V_{i}-J_{i})(\cos q_{i}\delta_{i}-
   \cos(k_{i}-k_{i}^\prime+q_{i})\delta_{i}),\label{spinpotencial}
   \end{gather}
where $i=x,y,z$ and $\delta_i=a,b,c$, the different matrix elements
involve the on site ($U$), nearest neighbour direct ($V_i$), exchange ($J_i$),
pair-hopping ($F_i$) and bond-charge ($C_i$) terms.
This interaction is able to support a variety of low temperature
phases\cite{Ozaki}, but we are only interested in unconventional
DW (whose gap depends on the perpendicular momentum)\cite{kiscikk,nagycikk}. The latter can be either UCDW or USDW 
depending on the strength of the exchange and pair-hopping integrals.
The single-particle electron thermal Green's function using Nambu's notation is\cite{parks,rickayzen}
\begin{equation}
G_\sigma({\bf k}, i\omega_n)=-\int_0^\beta d\tau\langle T_\tau \Psi_\sigma({\bf k}, \tau)
\Psi_\sigma^+({\bf k},0)\rangle_He^{i\omega_n\tau} ,
\end{equation}
where the Green's function is chosen to be diagonal in spin indices and the momentum space is divided into $\bf k$ and $\bf k-Q$ spaces (left- and right-going 
electrons) by
 introducing the spinors:
\begin{equation}
\Psi_\sigma({\bf k}, \tau)=
\left( \begin{array}{c}
         a_{\bf k, \sigma}(\tau) \\
         a_{\bf k-Q, \sigma}(\tau)
         \end{array}
 \right),
\end{equation}
$\omega_n$ is the Matsubara frequency, ${\bf Q}=(2k_F,\pi/b,\pi/c)$ is the best nesting vector. 
The inverse of the above Green's function is obtained as
\begin{equation}
G_\sigma^{-1}({\bf k}, i\omega_n)=i\omega_n-\xi({\bf k})\rho_3-
\Delta_\sigma({\bf k})\rho_1,\label{Green0}
\end{equation}
where $\rho_i$ ($i=1,2,3$) are the Pauli matrices acting on momentum space, 
$\Delta_\sigma(\bf k)$ satisfies the self-consistent equation:
\begin{equation}
\Delta_\sigma({\bf k})=\frac 1 V \sum_{\bf k^\prime,\sigma^\prime}\overline{\tilde{V}
({\bf k^\prime,k,Q},\sigma,\sigma^\prime)}\langle a_{\bf k^\prime,\sigma^\prime}^+a_{\bf k^\prime+Q,\sigma}\rangle
.
\end{equation}
In order to describe USDW, we assume $\Delta$ as an odd function of the spin ($\Delta_\sigma=-\Delta_{-\sigma}$).
Assuming $\Delta_\sigma$ to be an even function of the spin, we would have UCDW.
From now on, we will drop the spin indices since they are irrelevant for
most of our discussion and most of our results applies to both unconventional charge and spin density waves. 
The spin indices will be reinserted wherever necessary.
With this, the gap equation reads as
\begin{equation}
\Delta({\bf l})=\frac 1V \sum_{\bf k}\overline{P({\bf k,l})}\frac{\Delta({\bf k})
\tanh(\beta E({\bf k})/2)}{2E({\bf K})},
\end{equation}
where $E({\bf k})=\sqrt{\xi({\bf k})^2+|\Delta({\bf k})|^2}$,
$\Delta({\bf k})=\Delta_\sigma({\bf k})$ and the kernel
of the in\-teg\-ral equation is diagonal on the basis of the leading harmonics
as\cite{nagycikk}
\begin{eqnarray}
\frac{P({\bf k,l})}{V}=\frac{P_{0}}{N}+\frac{P_{1}}{N}\cos(k_{y}b)\cos(l_{y}b)
+\frac{P_{2}}{N}\sin(k_{y}b)\sin(l_{y}b)
+\frac{P_{3}}{N}\cos(k_{z}c)\cos(l_{z}c)+\frac{P_{4}}{N}\sin(k_{z}c)\sin(l_{z}c).
\end{eqnarray}
The $P_i$ coefficients are linear combinations of the interaction matrix
elements. As a consequence of the general
form of the kernel, the gap will be of the form
\begin{equation}
\Delta({\bf l})=\Delta_{0}+\Delta_{1}\cos(l_{y}b)+
\Delta_{2}\sin(l_{y}b)+\Delta_{3}\cos(l_{z}c)+\Delta_{4}\sin(l_{z}c).
\end{equation}
From now on we assume that only one kind of gap among the five possible
candidates, whose transition temperature is the highest, opens and 
persist all the way down to zero temperature. For example we find that USDW
is stable with respect to UCDW if $J_y\mp F_y>0$, where the upper (lower)
sign refers to a $k_y$ dependent gap function of cosine (sine)\cite{nagycikk}. The thermodynamic and
transport properties of such a system has been worked out in Ref. 
\onlinecite{nagycikk}.
In the followings we shall discuss the effect of impurities on UDW and 
determine the behaviour of the basic physical quantities.
The interaction of the electrons with nonmagnetic impurities is described by the
Hamiltonian:
\begin{eqnarray}
H_1=\frac1V \sum_{{\bf k,q}, \sigma, j}e^{-i\bf{qR}_j}\Psi_\sigma^+({\bf k+q})U({\bf R}_j)
\Psi_\sigma({\bf k}),\\
U({\bf R}_j)=\left( \begin{array}{cc}
         U(0) & U({\bf Q})e^{-i{\bf QR}_j} \\
         \overline{U({\bf Q})}e^{i{\bf QR}_j} & U(0)
         \end{array}
 \right),\label{impurity}
\end{eqnarray}
${\bf R}_j$ is the position of the $j$-th impurity atom. The explicit wavevector dependence of the matrix
elements\cite{haran,rapid} is neglected since no important changes are expected from it.
The usual method of treating the impurities is to average over their
position in real space, and step into the wavevector space
afterwards\cite{rickayzen,szummad-wave}. Instead, we follow a rather unorthodox way: working in the 
Fourier space and averaging when needed.
 It is clear from the exponential prefactor in $H_1$, that
only diagrams containing impurity scattering with momentum conservation at
each impurity atom have finite expectation value after averaging over the position of
the impurities, and translational invariance is regained. 
\begin{figure}[h!]
\begin{fmffile}{graph1}
\begin{eqnarray}
 \parbox{10mm}{\begin{fmfgraph}(20,40)\fmfsurroundn{e}{4}
  \fmf{phantom}{e1,v}
  \fmf{dashes}{e2,v}
  \fmfv{decor.shape=cross,decor.size=4thick}{e2}
  \fmf{phantom}{v,e3}
  \fmf{phantom}{v,e4}\end{fmfgraph}} \quad + \quad
\parbox{30mm}{\begin{fmfgraph}(60,40)
\fmfsurroundn{e}{4}
  \fmf{fermion}{e1,e3}
  \fmf{dashes}{e2,e1}
  \fmf{dashes}{e2,e3}
  \fmfv{decor.shape=cross,decor.size=4thick}{e2}
  \fmf{phantom}{e1,e4}
  \fmf{phantom}{e3,e4}\end{fmfgraph}} \quad + \quad
\parbox{30mm}{\begin{fmfgraph}(60,40)
\fmfsurroundn{e}{4}
  \fmf{fermion}{e1,v}
  \fmf{fermion}{v,e3}
  \fmffreeze
  \fmf{dashes}{e2,e1}
  \fmf{dashes}{e2,v}
  \fmf{dashes}{e2,e3} 
  \fmfv{decor.shape=cross,decor.size=4thick}{e2}
  \fmf{phantom}{e1,e4}
  \fmf{phantom}{v,e4}
  \fmf{phantom}{e3,e4}\end{fmfgraph}} \quad + \quad \dots
\nonumber
\end{eqnarray}
\end{fmffile}
\vspace*{-10mm}
\caption{Self energy corrections due to impurity scattering. The solid line
denotes the electron while the dashed line is for the electron-impurity 
interaction. Dashed lines coming from the same cross represent successive 
scattering of the electron on the same impurity. \label{fig:graph}}
\end{figure}
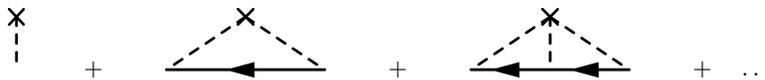

This method can
be extended into any order of impurity scattering as we will demonstrate it
in the followings. As to the diagrams, we will take into account only
non-crossing, ladder type diagrams\cite{rickayzen,klasszikus, abrikosov1}. Recently the applicability of this
approximation in 2 dimensions has been questioned and a new method has been invented in
order to consider all types of diagrams (i.e. those with crossing impurity
lines)\cite{Pepin}. As a result, a novel type of
$n_i/|2\omega|(\ln^2|\omega/\Delta|+(\pi/2)^2)$ additional density of states
was found, which could not be obtained by the non-crossing approximation,
$n_i$ is the impurity concentration. But
for 1 and 3 dimensional systems the usual technique looks sufficient\cite{suzumura,roshen}. 
To start with, we will evaluate the self e\-ner\-gy corrections caused by
Eq. (\ref{impurity}) at every order. This can be visualized in Fig. 
\ref{fig:graph} and is
given by 
\begin{gather}
\Sigma_{\bf R}({\bf k},i\omega_n)=\Sigma_{\bf R}(i\omega_n)=n_i\left(U({\bf R})+U({\bf R})\int\frac{d^3p}
{{2\pi}^3}G({\bf p},i\omega_n)U({\bf R})\right.+\nonumber \\
\left.+U({\bf R})\int\frac{d^3p}{{2\pi}^3}G({\bf p},i\omega_n)U({\bf R})
\int\frac{d^3p^\prime}{{2\pi}^3}G({\bf p}^\prime,i\omega_n)U({\bf
R})+\dots\right)=n_i U({\bf R})
+U({\bf R})\int\frac{d^3p}{{2\pi}^3}G({\bf p},i\omega_n)\Sigma_{\bf R}(i\omega_n),\label{sajatenergia}
\end{gather}
where the self energy correction turns out to be momentum independent and
the $\bf R$ index in $\Sigma_{\bf R}(i\omega_n)$ means the position of an impurity over which the average
will be taken in the followings.
Eq. (\ref{sajatenergia}) can be solved easily, and the result is:
\begin{eqnarray}
\Sigma_{\bf R}(i\omega_n)=\left( \begin{array}{cc}
         U_1-g & U_2e^{-i\bf QR}+f \\
         \overline{U}_2e^{i{\bf QR}}+\overline{f} & U_1-g
         \end{array}
 \right)
\frac{n_i}{(U_1-g)^2-|f|^2-|U_2|^2-(U_2\overline{f}e^{-i\bf
         QR}+\overline{U}_2f e^{i\bf QR})},\label{sigma}
\end{eqnarray}
where $U_1=U(0)/(U(0)^2-|U({\bf Q})|^2)$ and $U_2=U({\bf
Q})/(U(0)^2-|U({\bf Q})|^2)$ and
\begin{eqnarray}
\int\frac{d^3p}{{2\pi}^3}G({\bf p},i\omega_n)=\left( \begin{array}{cc}
         g & f \\
         \overline{f} & g
         \end{array}
 \right).
\end{eqnarray}
Expanding $\Sigma_{\bf R}(i\omega_n)$ in powers of the exponential terms in the denominator of
Eq. (\ref{sigma}), the space average can be performed and the self energy
matrix is obtained as
\begin{gather}
\Sigma(i\omega_n)=\left( \begin{array}{cc}
         \Sigma_1(i\omega_n) & \Sigma_2(i\omega_n) \\
         \Sigma_3(i\omega_n) & \Sigma_1(i\omega_n)
         \end{array}
 \right)
\end{gather} and its matrix elements are given by
\begin{gather}
\Sigma_1(i\omega_n)=n_i\frac{U(0)-g(U(0)^2-|U({\bf
         Q})|^2)}{\sqrt{D^2-4|U({\bf Q})f|^2}},\\
\Sigma_2(i\omega_n)=n_i\frac{f}{\sqrt{D^2-4|U({\bf
Q})f|^2}}\left(U(0)^2-|U({\bf Q})|^2 +\frac{2|U({\bf
Q})|^2}{D+\sqrt{D^2-4|U({\bf Q})f|^2}}\right),\\
\Sigma_3(i\omega_n)f=\Sigma_2(i\omega_n)\overline{f},
\end{gather}
 where $D=1-2gU(0)+(g^2-|f|^2)(U(0)^2-|U({\bf Q})|^2)$. This
result is valid only for a certain range of parameters due to the
expansion. On the other hand, one can deduce an expression involving the
different matrix elements of the self energy where the average can be performed
rigorously:
\begin{eqnarray}
\Sigma_1(1-2U(0)g+(g^2+|f|^2)(U(0)^2-|U({\bf Q})|^2))
-(\Sigma_2\overline{f}+\Sigma_3f+n_i)(U(0)-g(U(0)^2-U({\bf Q})^2)=0,
\end{eqnarray}
and this equation is satisfied with the previously obtained $\Sigma_1$,
$\Sigma_2$ and $\Sigma_3$ even outside of the validity range of the expansion. Of course this
cannot be regarded as a proof but we can trust in the usefulness of
this calculation outside the validity range. Moreover in a normal metal this result gives back the
known result\cite{klasszikus}. These formulas apply also to superconductors with minor
change ($U({\bf R})=U(0)\rho_1$), and the self energies in the Born and
unitary limit are obtained correctly\cite{parks,skalski,kadanoff,ambegaokar,impurd-wave}.
We treat our UDW system in the Born scattering limit since conventional DWs
are commonly investigated in this limit\cite{roshen}. The interaction gives rise to the self energy,
which is in the Born-approximation (considering only the lowest order terms):
\begin{gather}
\Sigma({\bf k},i\omega_n)=\frac{n_i}{V}\sum_{\bf q}\frac1N\sum_{\bf R} U({\bf R})
G({\bf k-q},i\omega_n)U({\bf R}),
\end{gather}
where the summation is the only remaining operation from averaging over the
impurity atoms. 
From this, one obtains for a DW
\begin{equation}
G({\bf k},i\omega_n)=-\frac{i\tilde\omega_n+\xi({\bf k})\rho_3+\tilde
\Delta_n({\bf k})\rho_1}{\tilde\omega_n^2+\xi({\bf k})^2+\tilde\Delta_n({\bf k})^2},
\end{equation}
where both the frequency and the gap are renormalized in the conventional 
case:
\begin{eqnarray}
\omega_n=\tilde\omega_n-\frac{\Gamma_1+\Gamma_2}{2}\frac{\tilde\omega_n}{
\sqrt{\tilde\omega_n^2+\tilde\Delta_n^2}},\\
\Delta=\tilde\Delta_n+\frac{\Gamma_1}{2}\frac{\tilde\Delta_n}{\sqrt{\tilde
\omega_n^2+\tilde\Delta_n^2}}.
\end{eqnarray}
$\Gamma_1=\pi n_i |U(0)|^2g(0)$ is the forward scattering,
$\Gamma_2=\pi n_i |U({\bf Q})|^2g(0)$ is the backward scattering
parameter. $n_i$ is the impurity concentration, $g(0)$ is the density of
states per spin in the metallic state.
As in other similar problems\cite{abrikosov1}, it is convenient to introduce the quantity
$u_n=\tilde\omega_n/\tilde\Delta_n$, which relates to physical quantities:
\begin{equation}
\omega_n=\Delta u_n\left(1-\alpha\frac{1}{\sqrt{u_n^2+1}}\right),
\end{equation}
$\Gamma=\Gamma_1+\frac{\Gamma_2}{2}$, $\alpha=\Gamma/\Delta$ is the 
pair-breaking parameter. As opposed to this, in unconventional DW 
self energy corrections from impurities does not 
renormalize the gap, only the Matsubara frequency:
\begin{eqnarray}
\omega_n&=&\tilde\omega_n-\frac{\Gamma_1+\Gamma_2}{\pi}\frac{\tilde\omega_n}
{\sqrt{\tilde\omega_n^2+\Delta^2}}K\left(\frac{\Delta}{\sqrt{\tilde\omega_n^2+\Delta^2}}\right),
\nonumber\\
\tilde\Delta_n({\bf k})&=&\Delta({\bf k})=\Delta\sin(bk_y)\textmd{ or 
}\Delta\cos(bk_y). 
\end{eqnarray}
This is written in a more useful dimensionless form:
\begin{equation}
\omega_n=\Delta u_n\left(1-\frac{2}{\pi}\frac{\alpha}{\sqrt{u_n^2+1}}
K\left(\frac{1}{\sqrt{u_n^2+1}}\right)\right),
\end{equation}
where $\Gamma=(\Gamma_1+\Gamma_2)/2$, $\alpha=\Gamma/\Delta$, $u_n=\tilde\omega_n/\Delta$ and $\Gamma_1$ and $\Gamma_2$ are the same 
quantities as in a conventional DW, $K(z)$ is the complete elliptic integral of
the first kind. 
Here the combination of the scattering rates is different from the conventional
DW's case due to the lack of renormalization of the order parameter. We choose
the Born scattering limit because this limit works very well for 
conventional DW. We believe that by neglecting the explicit wavevector dependence of the impurity
matrix elements, we made a useful approximation as far as the character of the
physics is concerned and we are able to capture the characteristic changes caused by
impurities. However in order to describe very fine, characteristic phenomena
to DW such as the threshold electric field\cite{fl,lr,viro1,viro2}, we cannot 
use simple s-wave
scatterers as it is shown in Refs. \onlinecite{rapid,tesla}. 

\section{Thermodynamics of impure UDW}

Since the thermodynamic properties
of a pure UDW are identical to those of a d-wave superconductor\cite{nagycikk,d-wave} and the impurity
effects on a conventional DW are similar to those in s-wave superconductors, we expect
very similar behaviours to those in a d-wave superconductor treated in the Born limit.
However, the main difference is that we distinguish two different scattering
processes (forward and backward scattering) while in the superconducting world
there is only one. Consequently the different combinations of the 
$\Gamma$'s are far from being trivial.
The gap equation is obtained as
\begin{equation}
1=\rho(0)T P_i\sum_n\left(E\left(\frac{1}{\sqrt{1+u_n^2}}\right)
\sqrt{1+u_n^2}-K\left(\frac{1}{\sqrt{1+u_n^2}}\right)\frac{u_n^2}{\sqrt{1+u_n^2}}\right),
\label{uncgapeq}
\end{equation}
where $E(z)$ is the complete elliptic integral of the second kind.
The change in the transition temperature is given by 
the Ab\-ri\-kos\-ov-Gor'kov formula:
\begin{equation}
-\ln\left(\frac{T_c}{T_{c_0}}\right)=\psi\left(\frac12+\rho\right)-
\psi\left(\frac12
\right),
\end{equation}
where $T_c$ and $T_{c_0}$ are the transition temperature of the impure 
and clean
system, respectively, $\rho=\Gamma/2\pi T_c$, $\psi(z)$ is the
digamma function.
 Note that this formula is also valid for 
any kind of unconventional superconductor in the presence of impurities 
considered either in Born or in resonant scattering limit\cite{szupravezetes}.
The critical impurity scattering rate is given by 
\begin{equation}
\Gamma_c=\frac{\pi T_{c_0}}{2\gamma}=\frac{\sqrt e \Delta_{00}}{4}
\end{equation}
The gap maximum is the same as the one of a d-wave SC in the Born limit\cite{epl2}:
\begin{equation}
\ln\frac{\Delta_{00}}{\Delta(0,\Gamma)}=\frac{8}{\pi^2}\frac{\Gamma}{\Delta}
\int_{C_0}^{\infty}(K-E)\left(E-K\frac{x^2}{1+x^2}\right)dx+2\langle
\sin^2 y \textmd{arsh}\frac{C_0}{\sin y}\rangle,
\end{equation}
where $\langle \dots \rangle$ means $\frac{1}{2\pi}\int_0^{2\pi}dy\dots$, the
argument of $K$ and $E$ reads as $\frac{1}{\sqrt{x^2+1}}$. $C_0$ is the value of
$u_n$ at zero frequency:
\begin{equation}
\sqrt{1+C_0^2}=\frac 2\pi \alpha K\left(\frac{1}{\sqrt{1+C_0^2}}\right),
\end{equation}
 vanishing as the impurity scattering parameter disappears like
$C_0=4\exp(-\pi/2\alpha)$, while for large $\alpha$: $C_0=\alpha$ as in Ref. \onlinecite{epl3}.
Close to $T_c$, $\Delta$ vanishes in a square-root manner as does usually in 
mean field treatments:
\begin{equation}
\Delta^2=8(2\pi T_c)^2 \frac{1-\rho\psi^{'} \left(\frac12 +\rho\right)}
{-\frac{3\psi^{''}\left(\frac12+\rho\right)}{2}-\rho\frac{\psi^{'''}
\left( \frac12 +\rho\right)}{3}} \left(1-\frac{T}{T_c}\right).
\end{equation}
Close to absolute zero, the following formula is obtained:
\begin{eqnarray}
\Delta(T)=\Delta(0)-\frac{\pi^2}{3}\frac{C_0}{\Gamma}\left(\frac K E
-1\right)\left(1-\frac{4}{\pi}\left(\frac{2\alpha}{\pi}\int_{C_0}^\infty(K-E)\times
\right.\right. \nonumber \\ 
\times \left.\left. \left(E-K\frac{x^2}{1+x^2}\right)dx+C_0\sqrt{1+C_0^2}\left(E-K\frac{C_0^2}{1+C_0^2}\right)\right)\right)^{-1}T^2,
\end{eqnarray}
where the $T^3$ decrease of the pure case turned into a faster $T^2$ one.
\begin{figure}[h]
\psfrag{x}[t][b][1][0]{$\Gamma/\Gamma_c$}
\psfrag{y}[b][t][1][0]{$\Delta(0,\Gamma)/\Delta_{00}$, $T_c/T_{c0}$ and 
$N(0,\Gamma)/N_0$}
\centering{\includegraphics[width=11cm,height=7cm]{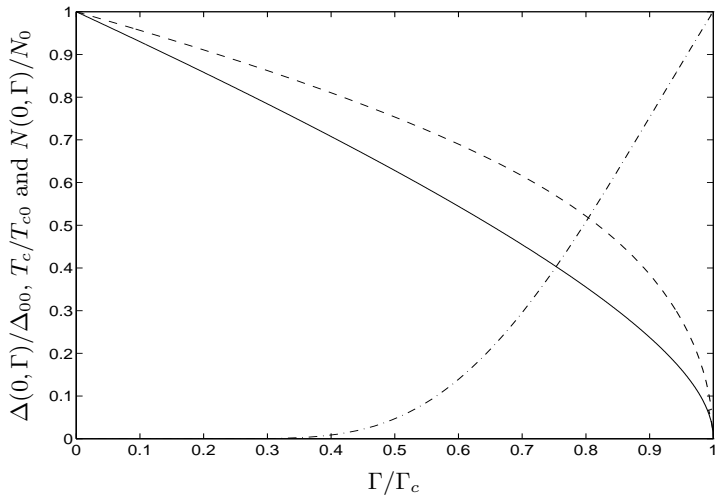}}
\caption{$\Delta(0,\Gamma)/\Delta_{00}$ (dashed line), $T_c/T_{c0}$ (solid
line) and $N(0,\Gamma)/N_0$ (dashed-dotted line) are shown as a function of
$\Gamma/\Gamma_c$ for an unconventional density wave.}
\label{fig:dtnbu}
\end{figure}

From this, one can assume that the effect of the impurity scattering in the
limit of low temperatures is to reduce the power-law exponent by one. As a
result we expect the exponent of temperature to be the same as those in a
conventional DW in the gapless region\cite{parks,roshen}. The analogy looks 
obvious since in
neither of these two systems there is a lower bound of the excitation
energy. The correspondence works only at low temperatures $T\ll T_c$ when
the only energy scale is the temperature.  
Now we derive expressions for the grand canonical potential
and for the specific heat. In doing this, we use the well-known relation involving
an integral over the coupling constant of the interaction\cite{klasszikus}:
\begin{equation}
\Omega-\Omega_0=\int_0^1\frac{d\lambda}{\lambda}\langle \lambda H_{int}\rangle,
\label{grandcanonical}
\end{equation}
where $H_{int}$ is the interaction causing the phase transition. This formula
gives us the thermodynamic potential difference between the normal and the DW phase.
Since we work on a grand canonical ensemble, the appropriate thermodynamic
potential at $T=0$ is obtained as:
\begin{eqnarray}
\Omega(0)=-N\rho(0)\left(\frac{\Delta^2}{4}-\frac2 \pi \Delta^2
C_0\sqrt{C_0^2+1}+\frac{\Gamma^2}{3}+\frac{2\Delta^3
C_0^3}{3\Gamma}-\frac{4\Gamma\Delta}{\pi^2}\int_{C_0}^\infty(K-E)\left(E-K\frac{x^2}{1+x^2}\right)dx\right).
\end{eqnarray}
At small $\Gamma$, the leading correction is the last integral, enhancing the
potential as in the normal SDW case. The low 
temperature specific heat reads as
\begin{equation}
C(T)=\frac{2\pi^2}{3}g(0)\frac{\Delta C_0}{\Gamma}T.
\end{equation}
This expression also reaches the normal state value with increasing
$\Gamma$. The specific heat jump is:
\begin{equation}
\Delta C(T)=\frac{16\pi^2g(0)T_c}{-\frac{3\psi^{''}\left(\frac12+
\rho\right)}
{2}-\rho\frac{\psi^{'''}\left(\frac12+\rho\right)}{3}}
\left(1-\rho\psi^{'}\left(\frac12+\rho\right)\right)^2.
\end{equation}
In Fig. \ref{fig:dtnbu} we show $\Delta(0,\Gamma)$ and $T_c$ as a function of the
scattering rate.

\section{Density of states in UDW}

\begin{figure}[h!]
\psfrag{x}[t][b][1][0]{$\omega/\Delta$}
\psfrag{y}[b][t][1][0]{$N(\omega)/g(0)$}
\centering{\includegraphics[width=11cm,height=7cm]{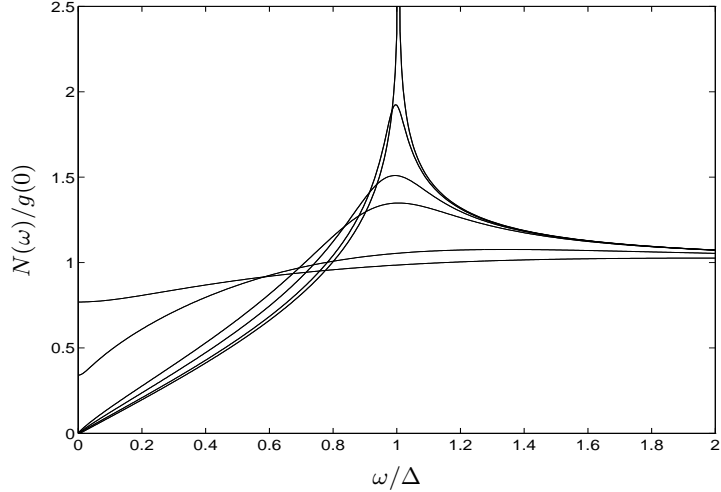}}
\caption{Density of states plotted as a function of the reduced energy for
different scattering amplitudes: $\alpha=0$, $0.01$, $0.05$, $0.1$, $0.5$ and
$1$ with peakposition at $\omega=\Delta$ from top to bottom.\label{fig:undos}}
\end{figure}

By use of the Green's function, the density of states per spin is given by:
\begin{equation}
N(\omega)=-\frac{1}{2\pi V}\sum_{\bf k}\textmd{ImTr}(G^R({\bf k},\omega))=
g(0)\frac1\alpha\textmd{Im}(u),
\end{equation}
where $u=iu_n(i\omega_n=\omega+i\delta)$. 
After some algebra, the low energy behaviour reads as:
\begin{equation}
N(\omega)=g(0)\frac{C_0\Delta}{\Gamma}\left(1+\frac{\pi^2}{8E^2}\left(\frac
K E +\frac{1}{C_0^2}-1\right)\left(\frac{\omega}{\Gamma}\right)^2\right).
\end{equation}
The residual density of states (i.e. the DOS at the Fermi energy) 
is finite at any finite
$\Gamma$, disappears exponentially as $\Gamma$ goes to zero, but takes the
normal state value as $\Gamma$ approaches to infinity. Since $N(0)$ is almost
zero for $\Gamma<0.5\Gamma_c$, we do not expect relevant changes in the static
quantities (such as the specific heat, the spin susceptibility at 
$T\rightarrow 0$) at low impurity concentrations. The notion "gapless"
makes no sense in this case since even in pure UDW the gap vanishes at the Fermi
energy leading to the possibility of arbitrary small energy excitations. 
At the value of the order parameter, the divergent peak of the pure system is
broadened and $N(\omega)$ is always finite as a
result of the impurities, shown in Fig. \ref{fig:undos}. Compared to the DOS of 
the conventional DW\cite{roshen}, the states below the gap maximum are 
filled in, and the peak at $\Delta$ is disappears more rapidly as $\alpha$ increases
than in the case of momentum independent gap.
At high energies it reaches the normal state value as:
\begin{equation}
N(\omega)=g(0)\left(1+\frac{\Delta^2}{4}\frac{\omega^2-\Gamma^2}
{(\omega^2+\Gamma^2)^2}\right).
\end{equation}
In Fig. \ref{fig:dtnbu}, we show the $\Gamma$ dependence of the residual density of 
states. 

\section{Density correlator}

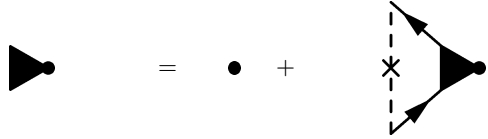
\begin{figure}[h!]
\begin{fmffile}{graph2}
\begin{eqnarray}
 \parbox{25mm}{\begin{fmfgraph}(80,50)
  \fmfleftn{a}{2}
   \fmfright{b}
   \fmfrpolyn{full}{e}{3}
  \fmfdot{e3}
 \fmf{phantom}{a1,e1}
 \fmf{phantom}{a2,e2}
 \fmf{phantom}{e3,b}
\end{fmfgraph}}      \quad = \quad
 \parbox{5mm}{\begin{fmfgraph}(5,25)
  \fmfleft{e1}
  \fmfright{e1}
  \fmfdot{e1}
\end{fmfgraph}} \quad + \quad
 \parbox{40mm}{\begin{fmfgraph}(80,50)
  \fmfleftn{a}{2}
  \fmfright{b}
  \fmfrpolyn{full}{e}{3}
  \fmfdot{e3}
 \fmf{fermion}{a1,e1}
 \fmf{fermion}{e2,a2}
 \fmf{dashes}{a1,v}
 \fmf{dashes}{v,a2}
 \fmfv{decor.shape=cross,decor.size=4thick}{v}
 \fmf{phantom}{e3,b}
\end{fmfgraph}}
\nonumber
\end{eqnarray}
\end{fmffile}
\caption{The vertex correction in the Born limit is shown. The dot is the vertex
function, the filled triangle represents the vertex correction due to impurity
scattering.\label{fig:vertex}}
\end{figure}

We turn our attention to the behaviour of the static, long wavelength density correlation
function\cite{roshen} using the the thermal Green's function:
\begin{equation}
\chi_0(T)=-\frac1\beta \sum_{{\bf p,k},\sigma,n}\textmd{Tr}\overline{(G({\bf p,k},i\omega_n)
G({\bf k,p},i\omega_n))}
\end{equation}
where the overline means averaging over the position of the impurity atoms.
This requires calculating the averaged Green's function and the vertex
corrections, since the average of the product of two Green's functions is
not equal to the product of the averaged Green's functions. In the following
we focus on the vertex corrections, $\Lambda({\bf p},i\omega_n)$.
With this, our equation becomes simpler:
\begin{equation}
\chi_0(T)=-\frac1\beta \sum_{{\bf p},\sigma,n}\textmd{Tr}(G({\bf p},i\omega_n)
\Lambda({\bf p},i\omega_n)G({\bf p},i\omega_n)).\label{nn1}
\end{equation}  
In the standard ladder approximation the vertex corrections are determined
by the integral equation:
\begin{equation}
\Lambda({\bf p},i\omega_n)=1+\frac{n_i}{V} \sum_{\bf q}\frac1N\sum_{\bf R}
U({\bf R})G({\bf q},i\omega_n)\Lambda({\bf q},i\omega_n)G({\bf q},i\omega_n)
U({\bf R}),\label{nn2}
\end{equation}
which is shown in diagrammatic lan\-gu\-a\-ge in Fig. \ref{fig:vertex}.
As\-su\-ming $\Lambda({\bf p},i\omega_n)=\Lambda(i\omega_n)$, and making the
following ansatz:
\begin{eqnarray} 
\Lambda(i\omega_n)=\left(
\begin{array}{cc}
\Lambda_1(i\omega_n) & \Lambda_2(i\omega_n) \\
\Lambda_2(i\omega_n) & \Lambda_1(i\omega_n)   
\end{array}
\right),\label{nn3}
\end{eqnarray}
the vertex corrections can be obtained:
\begin{eqnarray}
\Lambda_1&=&\left(1-\frac 2\pi \alpha\frac{K-E}{\sqrt{1+u_n^2}}\right)^{-1}\\
\Lambda_2&=&0. 
\end{eqnarray}
Substituting this to Eq. (\ref{nn1}), the susceptibility reads as:
\begin{equation}
\chi_0(T)=2g(0)\left(1-\dfrac{2}{\Delta\beta}\sum_n\dfrac{\dfrac{K-E}{\sqrt{u_n^2+1}
}}{1-\alpha\dfrac{K-E}{\sqrt{u_n^2+1}}}\right).
\end{equation}
At zero temperature it equals to the total density of states at the Fermi 
surface:
\begin{equation}
\chi_0(0)=2g(0)\frac{C_0\Delta}{\Gamma}.
\end{equation}
In the low temperature limit we obtain:
\begin{equation}
\chi_0(T)=2g(0)\frac{C_0}{\Gamma}\left(\Delta(T)+\frac{\pi^4}{24E^2}\left(\frac
K E -1+\frac{1}{C_0^2}\right)\left(\frac{T}{\Gamma}\right)^2\right).
\end{equation} 
Close to $T_c$, a similar expression to the normal SDW describes the
susceptibility:
\begin{equation}
\chi_0(T)=2g(0)\left(1+
\frac{2\psi^{''}\left(\frac12+\rho\right)\left(1-\rho\psi^{'}\left(\frac12+
\rho\right)\right)}{-\frac{3\psi^{''}\left(\frac12+\rho\right)}{2}-\rho\frac{\psi^{'''}\left(\frac12+\rho\right)}{3}}
\left(1-\frac{T}{T_c}\right)\right).
\end{equation} 
We refrain from the evaluation of the $\bf Q$-th Fourier component of the
density correlation function because in UDW this is not the quantity which
signals the phase transition, RPA corrections will not lead
to divergence, since the dominant unconventional channel does
not couple to charge or spin density\cite{nagycikk}. In conventional CDW or SDW, the 
$\bf Q$-th Fourier component of the charge density or the spin density turned
out to be the order parameter of the phase transition, respectively. As opposed
to this, in the unconventional scenario, the following phases and related order
parameters are found:
\vspace*{2mm}

\begin{center}
\begin{tabular}{|c|c|c|} \hline
  phase & gap & order parameter: the $\bf Q$-th \\
        &     & Fourier component of the \\  \hline 
   UCDW    & $\Delta\cos(bk_y)$ & electric current density \\ \cline{1-3}
   UCDW    & $\Delta\sin(bk_y)$ & kinetic energy density \\ \hline
  USDW & $\Delta\cos(bk_y)$ & spin current density \\ \cline{1-3}
  USDW     & $\Delta\sin(bk_y)$ & spin kinetic energy density \\ \hline
\end{tabular}
\end{center}
\vspace*{2mm}

These phases are already known as
orbital antiferromagnet\cite{Ners1}, bond-order wave\cite{Ozaki}, spin nematic 
state\cite{Ners2} and axial spin bond-order wave\cite{Ozaki}, respectively in the 
context of the two dimensional Hubbard model. Generally these order parameters can be
called
as the effective charge or spin density\cite{nagycikk}. 
The autocorrelation function of the above quantities will be divergent at
$T_c$ in the corresponding phase, because
these are the relevant quantities from the phase transition's point.

\section{Optical conductivity}

\begin{figure}[h!]
\psfrag{x}[t][b][1][0]{$\omega/\Delta_{00}$}
\psfrag{y}[b][t][1][0]{Re$\sigma_{yy}^{cos}(\omega)4\Delta_{00}/e^2g(0)v_y^2$}
\centering{\includegraphics[width=11cm,height=7cm]{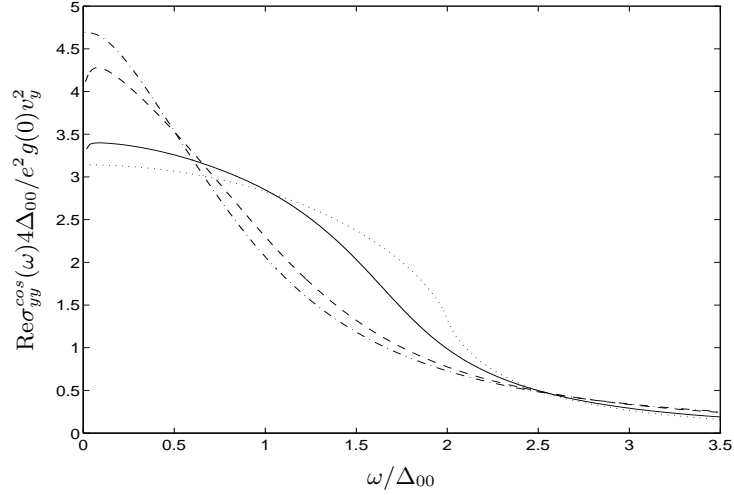}}
\caption{Real part of the electric conductivity in the $y$ direction for
$\Delta({\bf k})=\Delta\cos(bk_y)$ is plotted as a function of the reduced energy for
different scattering amplitudes: $\alpha=0$ (dotted line), $0.1$ (solid line),
$0.5$ (dashed line) and $1$ (dashed-dotted line), $\Gamma_1=2\Gamma_2$.
\label{fig:vyyc}}
\end{figure}

The
optical conductivity contains relevant informations about the possible
excitation of a system. Since in real materials impurities are always
present, the evaluation of the optical conductivity in impure systems is of
prime importance.  
As it is known, the electrical conductivity of a conventional DW is divided into
a pair-breaking (interband) and a normal (intraband) contribution\cite{epl1}. 
Hence a Lorentzian like normal contribution appears at all the frequencies, 
while the pair-breaking term is zero as long as $\omega<2\Delta$. This separation can be done in the
unconventional case, although here both processes contribute to all
frequencies due to the finite density of states at the Fermi energy.
Introducing two notations:
\begin{gather}
I_n(\omega)=\int_0^\infty \left(\tanh\frac{\beta(x+\omega)}{2}-\tanh\frac{\beta
x}{2}\right)\textmd{Re}(F(u(\omega+x),u(x))-F(u(\omega+x),\overline{u(x)}))dx,\\
I_{pb}(\omega)=\int_0^\omega
\tanh\frac{\beta(\omega-x)}{2}\textmd{Re}(F(u(\omega-x),-\overline{u(x)})-F(u(\omega-x),-u(x)))dx,
\end{gather}
the conductivity is given by:
\begin{equation}
\textmd{Re}\sigma_{aa}=-e^2g(0)v_a^2\frac{4}{\Delta \pi}
\frac{I_n(\omega)+I_{pb}(\omega)}{\omega},
\end{equation}
where $v_x=v_F$, $v_y=\sqrt 2 bt_b$ and $v_z=\sqrt 2 ct_c$. 
The different $F(u,{u'})$ functions and the dc conductivities are commented 
in the followings:
\vspace{2mm}

i.\hspace*{8mm} $\Delta({\bf k})=\Delta\cos(k_yb)$, $a=y$:
\begin{gather}
F(u,{u'})=\frac{1}{{u'}^2-u^2}\left[\sqrt{1-{u'}^2}\left(E'\left(-u{u'}-\frac23+\frac{{u'}^2}{3}\right)+K'\left(u{u'}-\frac{{u'}^2}{3}\right)\right)\right.+\nonumber
\\
\left.+\sqrt{1-u^2}\left(E\left(u{u'}+\frac23-\frac{u^2}{3}\right)+K\left(-u{u'}+\frac{u^2}{3}\right)\right)\right]
\end{gather}
This is the simplest case, the vertex corrections vanish due to the
mismatch of wavevector dependence of the velocity and the gap. As the
scattering strength enhances, it becomes the dominant energy scale and the curves take more and more the
form of a Lorentzian as it can readily be checked in Fig. \ref{fig:vyyc}. The DC conductivity is calculated at $T=0$:
\begin{equation}
\sigma_{yy}^{dc,cos}=e^2g(0)v_y^2\frac{4}{\Delta\pi}\left(E\sqrt{1+C_0^2}
-\frac{\pi C_0^2}{2\alpha}\right).
\end{equation}

\begin{figure}[h!]
\psfrag{x}[t][b][1][0]{$\omega/\Delta_{00}$}
\psfrag{y}[b][t][1][0]{Re$\sigma_{yy}^{sin}(\omega)4\Delta_{00}/e^2g(0)v_y^2$}
\centering{\includegraphics[width=11cm,height=7cm]{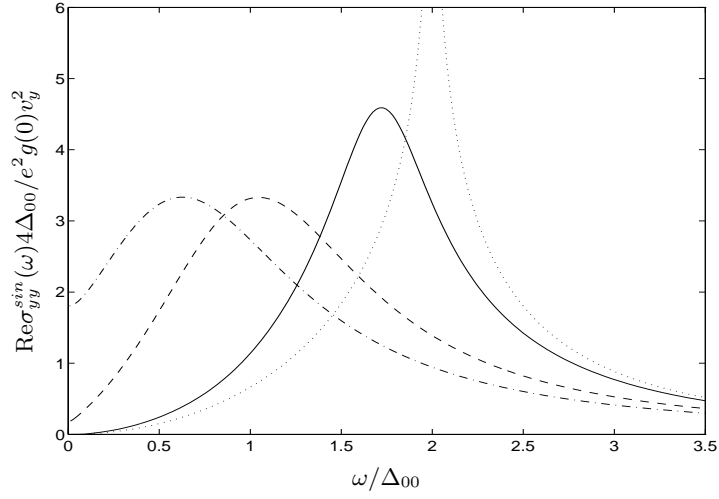}}
\caption{Real part of the electric conductivity in the $y$ direction for
$\Delta({\bf k})=\Delta\sin(bk_y)$ is plotted as a function of the reduced energy for
different scattering amplitudes: $\alpha=0$ (dotted line), $0.1$ (solid line),
$0.5$ (dashed line) and $1$ (dashed-dotted line), $\Gamma_1=2\Gamma_2$.
\label{fig:vyys}}
\end{figure}

\begin{figure}[h!]
\psfrag{x}[t][b][1][0]{$\omega/\Delta_{00}$}
\psfrag{y}[b][t][1][0]{Re$\sigma_{zz}(\omega)2\Delta_{00}/e^2g(0)v_z^2$}
\centering{\includegraphics[width=11cm,height=7cm]{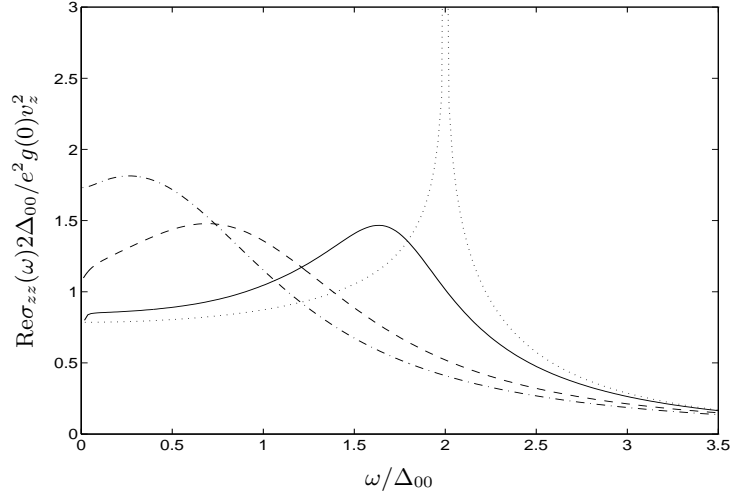}}
\caption{Real part of the electric conductivity in the $z$ direction is plotted
 as a function of the reduced energy for
different scattering amplitudes: $\alpha=0$ (dotted line), $0.1$ (solid line),
$0.5$ (dashed line) and $1$ (dashed-dotted line), $\Gamma_1=2\Gamma_2$.
\label{fig:vzz}}
\end{figure}

\vspace{2mm}

ii.\hspace*{8mm} $\Delta({\bf k})=\Delta\sin(k_yb)$, $a=y$:
\begin{gather}
F(u,{u'})=\frac{1}{{u'}^2-u^2}\left[\sqrt{1-u^2}E\left(-u{u'}+\frac43+\frac{u^2}{3}\right)
-\sqrt{1-{u'}^2}E'\left(-u{u'}+\frac43+\frac{{u'}^2}{3}\right)-\right.\nonumber \\
\left.
-\frac{{u'}^2}{\sqrt{1-{u'}^2}}K'\left(-u{u'}+\frac23+\frac{{u'}^2}{3}\right)+\frac{u^2}{\sqrt{1-u^2}}K\left(-u{u'}+\frac23+\frac{u^2}{3}\right)\right]+\nonumber
\\
+\dfrac{\Gamma_1}{\Delta\pi}\dfrac{1}{(u+{u'})^2}\dfrac{\left(E'\sqrt{1-{u'}^2}-E\sqrt{1-u^2}+
\dfrac{{u'}^2}{\sqrt{1-{u'}^2}}K'-\dfrac{u^2}{\sqrt{1-u^2}}K\right)^2}{1+\dfrac{\Gamma_1}{\Delta\pi}
\dfrac{1}{u+{u'}}\left(\dfrac{{u'}}{\sqrt{1-{u'}^2}}K'+\dfrac{u}{\sqrt{1-u^2}}K\right)},
\end{gather}
the third row of the equation comes from the vertex corrections. As
$\Gamma$ increases, the peak at $2\Delta$ is broadened and moves closer to zero frequency.
The DC conductivity is obtained at $T=0$:
\begin{equation}
\sigma_{yy}^{dc,sin}=4e^2g(0)v_y^2\frac{C_0^2(K-E)}{\Delta\pi\sqrt{C_0^2+1}+\Gamma_1(K-E)},
\end{equation}
where the second term in the denominator is clearly the effect of the
vertex corrections. The conductivity is shown in Fig. \ref{fig:vyys}. 
\vspace{2mm}

iii.\hspace*{8mm} $\Delta({\bf k})=\Delta\sin(k_yb)$ or $\Delta\cos(k_yb)$, $a=z$:
\begin{gather}
F(u,{u'})=\frac{1}{2({u'}^2-u^2)}\left(2\sqrt{1-u^2}E-2\sqrt{1-{u'}^2}E'
+K'\frac{{u'}(u-{u'})}{\sqrt{1-{u'}^2}}+K\frac{u(u-{u'})}{\sqrt{1-u^2}}\right),
\end{gather}
the vertex corrections vanish because the velocity depends on different perpendicular
wavevector component ($k_z$) than the gap ($k_y$). As $\Gamma$ increases, 
the peak at $2\Delta$ is broadened and moves closer to zero frequency.
The DC conductivity is obtained at $T=0$ as
\begin{equation}
\sigma_{zz}^{dc}=2e^2g(0)v_z^2\frac{E}{\Delta\pi\sqrt{C_0^2+1}},
\end{equation} 
The optical
conductivity is usually the same in the $x$ and $z$ direction apart from constant
factors, since the velocity in these directions does not interfere with the gap.
But in the presence of impurities this general relation does not hold any more
due to the presence of different vertex corrections. A very similar breakdown
of equality
is found in the relation between the static spin susceptibility and the condensate
density ($\rho_s=1-\chi_0/\chi_n$), which are not related to each other 
if impurity scattering is considered\cite{szummad-wave,impurd-wave}. 
The conductivity is shown in Fig. \ref{fig:vzz}. 
\vspace{2mm}

\begin{figure}[h!]
\psfrag{x}[t][b][1][0]{$\omega/\Delta_{00}$}
\psfrag{y}[b][t][1][0]{Re$\sigma_{xx}(\omega)2\Delta_{00}/e^2g(0)v_F^2$}
\centering{\includegraphics[width=11cm,height=7cm]{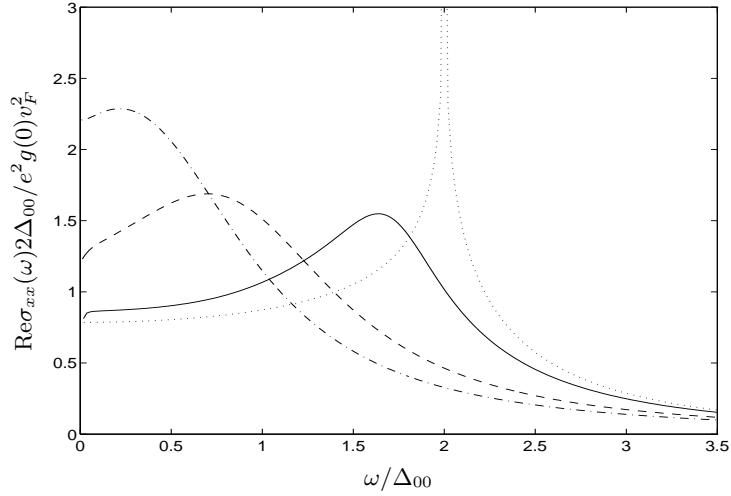}}
\caption{Real part of the electric conductivity in the chain direction 
is plotted as a function of the reduced energy for
different scattering amplitudes: $\alpha=0$ (dotted line), $0.1$ (solid line),
$0.5$ (dashed line) and $1$ (dashed-dotted line), $\Gamma_1=2\Gamma_2$. \label{fig:vxx}}
\end{figure}

For the sake of completeness we present the result for the quasiparticle
part of the conductivity
in the chain direction keeping in mind that collective modes also appear in
this direction.

iv.\hspace*{8mm} $\Delta({\bf k})=\Delta\sin(k_yb)$ or $\Delta\cos(K_yb)$, $a=x$:
\begin{gather}
F(u,{u'})=\frac{\pi\Delta}{2(\Gamma_1-\Gamma_2)}\left(\left(1-\frac{\Gamma_1-\Gamma_2}{\Delta\pi}
\left(-K\frac{u({u'}-u)}{\sqrt{1-u^2}}-K'\frac{{u'}({u'}-u)}{1-{u'}^2}
+2E\sqrt{1-u^2}-2E'\sqrt{1-{u'}^2}\right)\right)^{-1}-1\right).
\end{gather}
This formula gives the quasiparticle part of the optical conductivity in the
chain direction, although collective modes also show up here significantly
 modifying the conductivity. 
The consideration of impurity scattering and collective modes
(even in the simplest random-phase approximation) together is a very difficult 
task to deal with\cite{nakane1,nakane2} and is beyond the scope of the present investigation. 
The DC conductivity is obtained at $T=0$:
\begin{equation}
\sigma_{xx}^{dc}=2e^2g(0)v_x^2\frac{E}{\Delta\pi\sqrt{C_0^2+1}-(\Gamma_1-\Gamma_2)E}.
\end{equation}
The conductivity seems to transfer more and more spectral weight to the zero
frequency peak with growing impurity scattering rate, transforming the curve into a
Lorentzian like one (Fig. \ref{fig:vxx}).

\begin{figure}[h!]
\psfrag{x}[t][b][1][0]{$\Gamma/\Gamma_c$}
\psfrag{y}[b][t][1][0]{$\sigma(0)\Delta_{00}/e^2 g(0) v_{y,z}^2$}
\centering{\includegraphics[width=11cm,height=7cm]{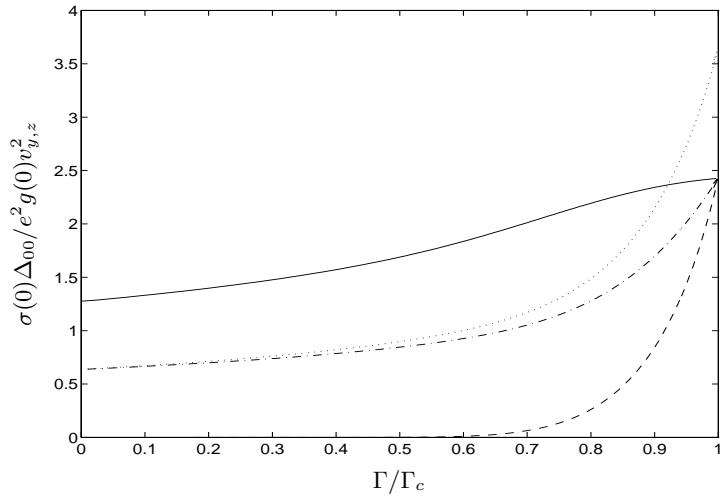}}
\caption{The dc conductivity plotted at $T=0$ as a function of the reduced 
scattering rate for $\Gamma_1=2\Gamma_2=4\Gamma/3$ for a cosinusoidal (sinusoidal)
gap in the $y$ direction: solid (dashed line), in the $z$ direction: 
dashed-dotted line and in the $x$ direction: dotted line. \label{fig:dc}}
\end{figure}

The dc conductivities are shown in Fig. \ref{fig:dc} at $T=0$ as a function
of the impurity scattering parameter. In the perpendicular direction, the dc
conductivities take the same value at the critical scattering parameter, while the
dc conductivity in the chain direction is exactly $3/2$ times larger as follows from Eq. 
(\ref{lorentzx}) and (\ref{lorentzy}) in the $\omega=0$ limit, 
if $\Gamma_1=2\Gamma_2$.
In spite of the similar
thermodynamics of d-wave SC\cite{epl2,epl3} and UDW, the transport properties of these two systems
are completely different due to the distinct coherence factors coming from
the different condensates. In a SC, there is
always a Dirac delta peak at zero frequency, and the stronger the impurity scattering, 
the larger the spectral weight of this peak transferred to the finite frequency part of 
the conductivity. In UDW, the Dirac delta contribution disappears as soon as any 
finite impurity concentration is present, and the areas under the different curves
are equal, but their form approaches those in the normal metal as $\Gamma$ enhances.

The normal state electric conductivities are given by the usual
Lorentzians:
\begin{eqnarray}
\textmd{Re}\sigma_{xx}(\omega)=e^2g(0)2v_F^2\frac{2\Gamma_2}{\omega^2+(2\Gamma_2)^2},
\label{lorentzx}\\
\textmd{Re}\sigma_{yy,zz}(\omega)=e^2g(0)2v_{y,z}^2\frac{\Gamma_1+\Gamma_2}{\omega^2+(\Gamma_1+\Gamma_2)^2}.
\label{lorentzy}
\end{eqnarray}
In the chain direction only backscattering can cause current damping as it is known
from transport theory, which is manifested in the absence of the forward scattering
parameter in Re$\sigma_{xx}(\omega)$.

\section{Conclusion}

We have studied the effect of nonmagnetic impurities in unconventional 
density waves. In this respect there is no difference
between USDW and UCDW due to the spin independence of the interaction with 
impurities.
In s-wave superconductors nonmagnetic impurities have no influence on the
thermodynamics of the system, while impure d-wave superconductors suffer important
changes. This is known as Anderson's theorem, but equivalent conclusion has
been reached independently by Abrikosov and Gor'kov. It says that if a
static perturbation does not break the time-reversal symmetry and does not
cause a long-range spatial variation of the order parameter, the
thermodynamic properties of the superconductor remain unchanged in the
presence of perturbation. As opposed to this, any kind of DW is destroyed in 
the presence of
impurities, although the identity of the thermodynamics of s-wave
superconductor to conventional DW and d-wave superconductor to unconventional
DW is well established without impurities. Impurities have a pair breaking
effect on the condensate, resulting in a universal formula between the
transition temperature and the scattering parameter, named after Abrikosov
and Gor'kov. It seems to be valid for superconductors with all kinds of
symmetries and now for density waves as well, independent of whether the
Born or the resonant scattering limit is taken. Since conventional DW were
studied in the Born limit, we found sufficient to use the same approximation
for the unconventional scenario. We have examined the system with the standard 
non-crossing
approximation, and calculated the self-energy corrections for infinite order
in the scattering potential, but only the lowest non-trivial correction was
retained for the Born limit. The thermodynamics of UDW were found to be 
very similar to the one in d-wave superconductors with nonmagnetic
impurities, but the existence of two different types of scattering
processes (forward and backward) was called for in the microscopic
theory. 
In unconventional DW, at any finite scattering strength the valley of the
density of states at the Fermi energy is filled in, leading to normal
electron like behaviours very close to absolute zero, but the reduced
density of states compared to the normal state bears the effect of the
condensate.
The order parameter does not get renormalized due to impurities because we
assumed s-wave scattering for simplicity. The specific heat increases 
linearly with temperature due to the finite density of states at the 
Fermi energy. Interestingly, impure UDW was found to be very similar to the 
gapless region of 
conventional DW very close to the critical scattering rate as long as the 
temperature exponents are concerned close to absolute zero because of the 
absence of any finite lower barrier of the excitation energy.

But at the transport properties all the
similarities ended. The optical conductivity in the chain direction is
dominated by the phason contribution, and incorporating the effect of
impurities in the theory is beyond the scope of this study. Instead we
concentrated on the perpendicular direction. 
In the optical conductivity, self
energy and vertex corrections were taken into account in the ladder type
non-crossing approximation. Depending on the symmetry of the order
parameter and the chosen direction, four qualitatively different curves are
deduced, although $\sigma_{xx}$ is certainly dressed by collective modes
due to coupling to the phason propagator. In the perpendicular directions, the possibility of low frequency excitations
rapidly increases, transferring increasing amount of spectral weight to
$\omega=0$. The dc conductivities at $T=0$ sharply differ from each other
hence they can help to provide one with decisive conclusion when comparing
these results to experimental data.

\begin{acknowledgments}
One of the authors (B. D.) greatfully
acknowledges the hospitality of the Max Planck Institute for the
Physics of Complex Systems, Dresden, where part of this work was done.
This work
was supported by the Hungarian National Research Fund under grant numbers
OTKA T032162 and T0374513, and by the Ministry of Education under grant 
number FKFP 0029/1999.
\end{acknowledgments}

\bibliographystyle{apsrev}
\bibliography{eth}
\end{document}